\documentclass[runningheads]{llncs}
\usepackage[T1]{fontenc}
\usepackage{graphicx}
\usepackage{hyperref}
\usepackage[nohyperlinks,nolist]{acronym}
\usepackage{subcaption}
\usepackage{rotating}
\usepackage{bbding}

\begin{document}
\title{An Analysis of the Impact of Psychological Factors and Techniques Across Different Types of Social Engineering}
\titlerunning{Social Engineering: Psychological Factors}
%
\author{Helin Omer\inst{1}\orcidID{0009-0004-0991-1617} \and Daniela Pöhn\inst{2}\orcidID{0000-0002-6373-3637}\Envelope}
\authorrunning{H. Omer and D. Pöhn}
%
\institute{Ludwig-Maximilians-Universität München, Munich, Germany \email{h.omer@campus.lmu.de} \and University of the Bundeswehr Munich, Neubiberg, Germany \email{daniela.poehn@unibw.de}}
\maketitle              
\begin{abstract}
Phishing is a well-known social engineering (SE) type used to trick individuals into revealing personal information or performing desired actions, like downloading and installing malware. Other SE types, like vishing and smishing, have emerged and are increasingly being used. As SE continues to successfully persuade victims into actions, the questions arise of which SE attack types are most effective for specific psychological factors (PFs) and, conversely, which PFs are most effective for particular attack types. To answer these questions, we conducted a laboratory study with n=12 participants, in which each participant was shown all 25 stimuli (five PFs and five SE types). The results of this exploratory study show that the most effective SE attack type for authority, trust, and greed was spear-phishing. The most successful combination of PF and attack type was spear-phishing using greed. The least successful combinations were pop-ups using authority, smishing using authority, and vishing using curiosity, each having had no success at all.

\end{abstract}

\section{Introduction}

 Human factors, including their behaviours, attitudes, and decision-making, influence how individuals manage cybersecurity issues and, ultimately, affect the effectiveness of cybersecurity measures. These human factors are exploited in \ac{SE} attacks by using \acp{PF} to trick individuals into revealing sensitive information or performing desired actions. Despite decades of innovations and improvements in cybersecurity technologies, humans remain the weakest link in organizational cybersecurity. The 2025 Verizon Data Breach Investigations Report~\cite{verizon} states that the human element is involved in around 60\,\% of all breaches. In a survey published in 2025, the SANS Institute~\cite{sans} reveals that 80\,\% of organizations rank \ac{SE} as the number one human-related risk, with phishing still leading. The survey further states that \ac{smishing} and \ac{vishing} attacks are growing in frequency and sophistication. As email filters have improved, attackers may focus more on other channels~\cite{apwg}.

 While many existing studies have examined individual aspects of human factors in cybersecurity~\cite{app12126042} or have focused on specific \ac{SE} attacks~\cite{longtchiInternetBasedSocialEngineering2024a}, especially phishing, few have analysed and compared different \acp{PF}, such as trust, authority, or curiosity, and \acp{PT}, which are strategies used by \ac{SE} attacks to exploit \acp{PF}, e.\,g., persuasion. Moreover, there is limited research combining \acp{PF} and \acp{PT} with different \ac{SE} attacks. By knowing the influence of the \acp{PF}/\acp{PT} on different \ac{SE} attacks, better awareness campaigns and other cybersecurity measures can be designed.
 
 This paper aims to bridge this gap by conducting a qualitative laboratory study. The study was conducted by exposing the participants (n=12) to different \ac{SE} attacks (n=5; phishing, spear-phishing, \ac{vishing}, \ac{smishing}, and pop-ups) and \acp{PF}/\acp{PT} (n=5; authority, fear, greed, trust, and curiosity), resulting in 25 stimuli. The answers were captured on a survey and through the think-aloud method. Our contribution is as follows: (1) comparison of different \acp{PF} and \acp{PT} for a given \ac{SE} attack; (2) comparison of different \ac{SE} attacks for a given \ac{PF}/\ac{PT}; (3) best and worst combinations.

The remainder of this paper is as follows: In Section~\ref{sec:background}, we provide the background on \ac{SE} and psychology. This is followed by a summary of related work in Section~\ref{sec:sota}. The study design is outlined in Section~\ref{sec:study-design}, followed by the presentation of the results in Section~\ref{sec:results}. The results are being discussed in Section~\ref{sec:discussion}, before we conclude the paper.

\section{Social Engineering and Psychology}
\label{sec:background}

\acp{PF} refer to individual characteristics that can be exploited by adapting the attacker's methods to an individual’s personality, aiming at that person’s psychological susceptibilities~\cite{longtchiCharacterizingEvolutionPsychological2025}. The Big Five Personality Traits~\cite{COBBCLARK201211} (openness, conscientiousness, extraversion, agreeableness, and neuroticism) is a model in psychology used to describe and assess human personality. Other areas that play a role are cognitive (e.\,g., dual-process theory~\cite{chaiken1999dual}) and emotion (e.\,g., empathy, sympathy, fear, and greed) psychology. Social psychology reflects on how individuals interact with others, focusing on behaviours related to connection, influence, and making or responding to requests. Cialdini~\cite{cialdiniSciencePersuasion2001} describes six principles of persuasion, namely reciprocation, consistency, social validation, liking, authority, and scarcity. Depending on the setting, workplace psychology with factors like stress, busyness, and habituation can play a role.

Longtchi et al.~\cite{longtchiInternetBasedSocialEngineering2024a} identified 16 \acp{PT}, which exploit various \acp{PF}: persuasion, attention grabbing, impersonation, pretexting, urgency, visual deception, priming, incentive and motivator, personalization, quid-pro-quo, foot in the door, and decoy effect. The authors conclude that fear, greed, authority, trust, and negligence are most exploited \acp{PF}.

\ac{SE} comprises methods of psychological manipulation that exploit human vulnerabilities to persuade individuals to take specific actions, like downloading and installing malware, or disclose confidential information, like usernames and passwords~\cite{pozoSocialEngineeringApplication2018}. There are different email-based attacks, such as generic mass phishing, spear-phishing, clone phishing, and whaling. Other channels are mobile phones (e.\,g., vishing, smishing, QR code phishing, or app spoofing), \ac{OSN} (e.\,g., honeytrap, catfishing, angler phishing, or fake profiles), websites (e.\,g., click-baiting, ad fraud, malvertising, and pop-up attacks), and the physical world (e.\,g., tailgating, baiting, dumpster diving, and shoulder surfing).

\section{Related Work}
\label{sec:sota}

Some publications generally discuss \ac{SE}. Mouton et al.~\cite{moutonSocialEngineeringAttack2014} provided a more detailed and structured model of \ac{SE} attacks by extending Mitnick's attack cycle.

Other publications focus on the \acp{PF} used in \ac{SE}. Longtchi et al.~\cite{longtchiInternetBasedSocialEngineering2024a} investigated why internet-based SE has remained successful despite existing defences by conducting an iterative literature review. The authors identified the gap that attackers exploit more psychological elements than defences addressed at that time. Bullée et al.~\cite{bulleeAnatomySocialEngineering2018} examined how persuasion principles are employed in SE attacks. The authors concluded that authority is being used most frequently.

Several studies linked personality traits and behaviour to susceptibility, highlighting the influence of conscientiousness and individual differences in information processing. Wall et al.~\cite{wallPersonalityProfilesPersuasion2019} investigated how personality profiles relate to susceptibility to persuasion. Butavicius et al.~\cite{butaviciusBreachingHumanFirewall2016} investigated how the three SE strategies of authority, scarcity, and social proof affect users’ judgments of email link safety. Halevi et al.~\cite{haleviPhishingPersonalityTraits2013} and Alseadoon et al.~\cite{alseadoonWhatInfluenceUsers2015} both investigated how user characteristics shape vulnerability to phishing attacks, whereas Halevi et al.~\cite{haleviSpearPhishingWildRealWorld2015} examined factors influencing the susceptibility to spear-phishing. Jun et al.~\cite{10.1145/3767320} conducted a questionnaire to understand the knowledge, behaviour, experience, and opinion on human factors in cybersecurity. The answers revealed a lack of comprehension of human factors and social engineering. Chrysanthou et al.~\cite{CHRYSANTHOU2024103780} provided a deep dive into large-scale phishing campaigns aimed at Meta's users. The authors find several poor password choices from the victims.

Longtchi and Xu~\cite{longtchiCharacterizingEvolutionPsychological2025} analysed psychological tactics and techniques in malicious emails using a real-world dataset. They found that certain tactics (e.\,g., fit and form) and techniques (e.\,g., attention grabbing) are used most often and often in combination. Burda et al.~\cite{10.1145/3635149} carried out a systematic literature review on the topic of \ac{SE}. The authors conclude that the experiments on \ac{SE} only partially reproduce real attacks and that the attack surface appears larger than the coverage provided by research. For example, attacks conveyed by media other than emails and websites are under-represented, according to the authors. Also, the number of \acp{PF}/\acp{PT} across studies is limited, with scarcity/urgency, authority, and liking being the most used ones.

Despite the number of publications, there is, to our knowledge, no study that examines the impact of \acp{PF} and \acp{PT} across different types of SE attacks.

\section{Study Design}
\label{sec:study-design}

The laboratory study employed a within-subject 5 x 5 design to examine how five specific \acp{PF} influence twelve participants' susceptibility to five different types of \ac{SE} attacks. The five \acp{PF} are curiosity, greed, trust, authority, and fear.
The five \ac{SE} attack types are phishing and spear-phishing emails, \ac{smishing}, \ac{vishing}, and pop-up attacks. This resulted in 25 stimuli, which were presented to each participant in a randomized order to minimize fatigue and order effects. Every \ac{PF} had their own different scenario. To ensure comparability within each \ac{PF}, the content was kept constant for the respective \ac{PF} and was presented only in different formats corresponding to the five attack types (e.\,g., in an email, SMS, voice message, or pop-up window). This approach ensured that the differences in participants' responses could be attributed to the manipulated variables rather than to content differences.
The \acp{PF} and \ac{SE} attack types were the two independent variables, while the participants' response intentions and verbalized reasoning were the main dependent variables.

\subsection{Recruitment}

A total of 12 participants were recruited through a combination of online and personal outreach strategies in Munich, Germany. They were required to be over 18 years of age. No compensation was given. The gender distribution was skewed towards females (nine female (75\,\%) and three males (33.3\,\%)). The ages of the participants ranged from 20 to 50 years, with a median of 22.5 years. Due to the described recruitment method, the sample may have been prone to self-selection bias.

\subsection{Procedure}
Each participant took part in an individual session lasting approximately 30-60 minutes. The study, including the stimuli and questions, was conducted in German, as this was the native language of the majority of the participants. Before starting with the main part of the study, participants were asked to pretend that their name is Max Mustermann\footnote{Equivalent to John Doe in Germany.}, while everything else about them would remain the same. Participants were also encouraged to use the think-aloud method for qualitative insights into cognitive and emotional responses regarding their decisions.

\subsubsection{Main Task} After each stimulus, participants were asked to answer the following questions on a tablet: 

\begin{itemize}
    \item ``Would you respond to the message/call?'' (yes/no)
    \item ``Why or why not?'' (open text field)
    \item ``How appealing did you find this message/call ?'' (1-4 Likert scale)
    \item ``How urgent did this message/call feel to you?'' (1-4 Likert scale)
\end{itemize}

\subsubsection{Post-Stimulus Questionnaire:}
After all 25 stimuli were shown, the post-stimulus questionnaire assessed information on participants' socio-demographics, technological affinity, experience, and their handling of unusual messages.

\subsubsection{Debriefing:} After completing the post-stimulus questionnaire, participants were asked a few reflective questions about their impressions and suspicions. They were then fully debriefed about the study's purpose and informed that all materials were simulated.

\subsection{Experimental Setup} 
The study was conducted in a laboratory environment to ensure consistent conditions for all participants. Each session took place in a testing room, with a tablet placed in front of the participant. All experimental materials, including the stimuli, stimulus-questions, and post-stimulus questionnaire, were presented on that tablet.

\subsubsection{PFs} For the \acp{PF}, the following impersonations and pretexts were used.

\begin{itemize}
\item \textbf{Authority:} A person impersonating the CFO of a company asked the recipient to urgently open an attached financial report and confirm its status the same day.
\item \textbf{Curiosity:} Google was impersonated and the used pretext was receiving access to a shared folder called ``Holiday party at the office''. 
\item \textbf{Greed:} The used pretext was receiving early access to a Black Friday deal from Amazon. 
\item \textbf{Fear} PayPal was impersonated and the pretext of unusual activity on the recipient's account was given, which allegedly resulted in the account being temporarily frozen. The recipient was asked to confirm their identity through a link.
\item \textbf{Trust:} Netflix was impersonated and the user was informed of updated account details. To view these details, the user was required to click the provided button. 
\end{itemize}

\subsubsection{SE} The following explains the design and creation of the attack examples.

\begin{itemize}
\item \textbf{Phishing and Spear-Phishing:} Screenshots of simulated phishing emails in an email inbox were presented. The simulated phishing emails were sent from the website canIPhish \cite{caniphishSecurityAwarenessTraining}. \item \textbf{Smishing:} SMS messages were created with the website iFake \cite{hessFakeTextMessage} and presented as screenshots in a smartphone layout. 
\item \textbf{Vishing:} For this attack type, the website ElevenLabs \cite{elevenlabsAIVoiceGenerator} was used to generate text-to-speech audio using one voice type for all audio to keep consistency among the different \ac{PF} examples and to avoid different answers based on the voice type. 
\item \textbf{Pop-Ups:} Screenshots of the Wikipedia main page as background with self-created pop-up windows were used. The background remained the same for all pop-up examples to ensure that different behavioural intentions could be attributed to the \ac{PF} instead of the background. 
\end{itemize}

\subsubsection{Material and Stimuli} 
\begin{figure}
    \centering
    \begin{subfigure}[b]{0.45\textwidth}
        \centering
        \includegraphics[width=0.95\linewidth]{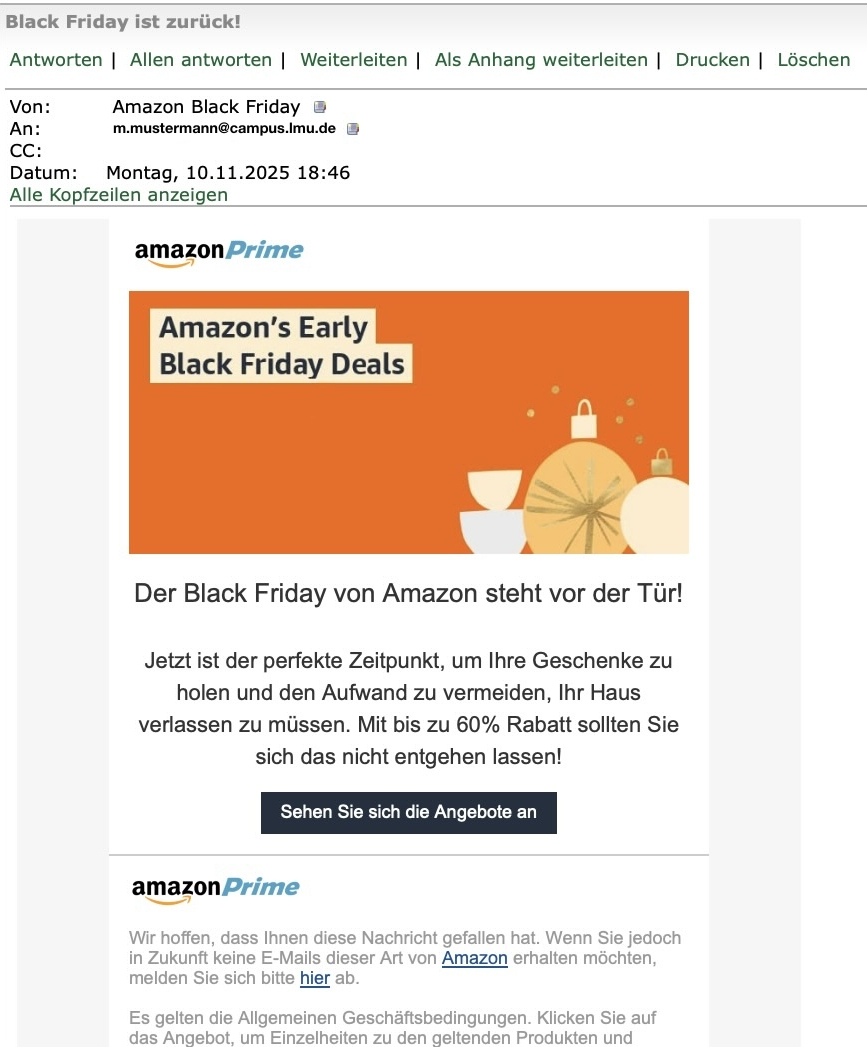}
        \caption{Phishing email using greed.}
        \label{fig:greed}
    \end{subfigure}
    \hfill
    \begin{subfigure}[b]{0.45\textwidth}
        \centering
        \includegraphics[width=0.95\linewidth]{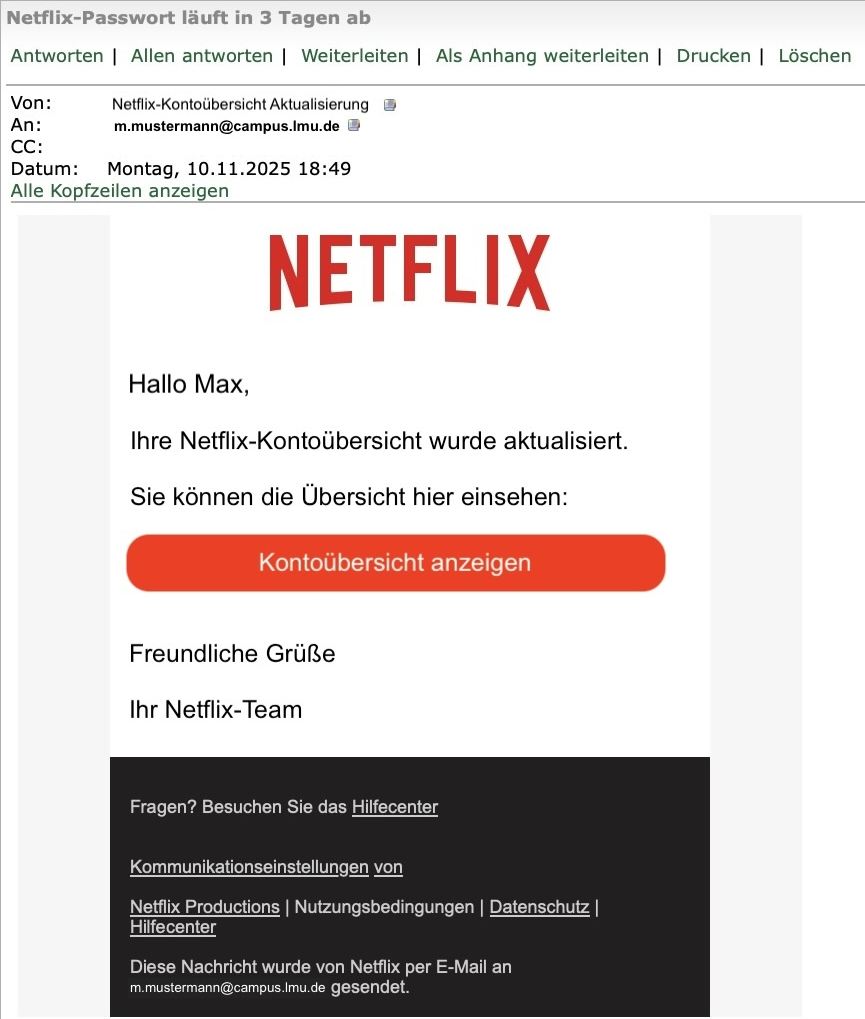}
        \caption{Spear-phishing email using trust.}
        \label{fig:trust}
    \end{subfigure}
    \caption{SE examples with different PFs.}
    \label{fig:examples}
\end{figure}

None of the materials were interactive or contained active links. The stimuli consisted of screenshots or short audio recordings designed to simulate realistic, yet harmless, examples of \ac{SE} attacks. Two examples can be seen in Figure~\ref{fig:examples}, with the combinations of phishing and greed (see Figure~\ref{fig:greed}) and spear-phishing and trust (see Figure~\ref{fig:trust}). Translated, Figure \ref{fig:greed} reads: ``Black Friday is just around the corner! Now is the perfect time to get your gifts and avoid the hassle of having to leave your home. With discounts of up to 60\,\%, you won't want to miss out! Check out the deals!''. Figure \ref{fig:trust} reads in English: ``Hello Max, your Netflix account details have been updated. You can look into your overview through this link: show account overview''.

\subsection{Measurements}
Both stimulus and post-stimulus questionnaires were used to assess quantitative and qualitative information.

\subsubsection{Quantitative Measures} The post-stimulus questionnaire assessed information on age, gender, study background, prior exposure to phishing, and participants' familiarity with digital means of communications and devices. Participants' behavioural intention to interact with the presented message was measured by their response to the question ``Would you respond to the message/call?'' (Yes/No). The questions: ``How appealing did you find
this message?'' and ``How urgent did this message feel to you?'' were asked to obtain
an estimation of the urgency and personal appeal to the participants.

\subsubsection{Qualitative Measures} Participants' verbalizations from the think-aloud protocol were audio-recorded and transcribed. Written explanations from the stimulus questionnaire (``Why or why not?'') were analysed qualitatively to identify reasoning patterns, such as risk awareness, cue utilization, emotional, or personal reactions.

\subsection{Limitations}
The small sample size and laboratory setting reduce the generalizability of the findings. Because the stimuli were presented as screenshots or audio and involved no real consequences, the ecological validity is limited. The \ac{SE} attacks can exploit multiple \acp{PF} across different \acp{PT}. This complexity was not addressed in the study, as this would have exceeded the scope of the thesis. Each \ac{PF} was represented by a single standardized scenario, which restricts the diversity in which each \ac{PF} can occur.

\section{Results}
\label{sec:results}

\subsection{Within-PF Design}

\subsubsection{Authority}
The most effective \ac{SE} type for authority was spear-phishing, with four out of twelve participants tending to interact with the phishing email (33\,\% yes-votes), followed by phishing. Vishing, \ac{smishing}, and pop-ups each had success rates lower than 10\,\%. The overall success rate of authority across the five \ac{SE} attacks was 13.3\,\%. When asked why participants would not interact with the stimulus, many answered that the stimulus was delivered on an inappropriate communication channel (in vishing attacks: \textit{``That could/should have been asked in an email.''}), that the design seemed illegitimate (\textit{``Does not seem legitimate''}), or that they were sceptical/distrusting (\textit{``Why do I have to download the file immediately?''}).

\subsubsection{Curiosity}
Curiosity was most successful in the spear-phishing and phishing scenarios, each having had a success rate of 50\,\%. Pop-ups, smishing, and vishing each had success rates lower than 10\,\%. The total number of yes-votes for curiosity in the five stimuli was 23.3\,\%, with spear-phishing and phishing each having a 42.8\,\% contribution to the total success rate. The most common reason for not interacting with stimuli that use curiosity included ``Scepticism / Distrust'' (30\,\% of all reasons for No-votes), followed by ``Not interested'' (23.9\,\% of reasons for No-votes). Participants who would interact with any stimuli using curiosity mostly reasoned their answer within the category ``Curiosity / Interest'' (57.1\,\% of reasons for Yes-votes), followed by ``Seems legitimate''.

\subsubsection{Fear}
\begin{figure}
    \centering
    \includegraphics[width=0.75\linewidth]{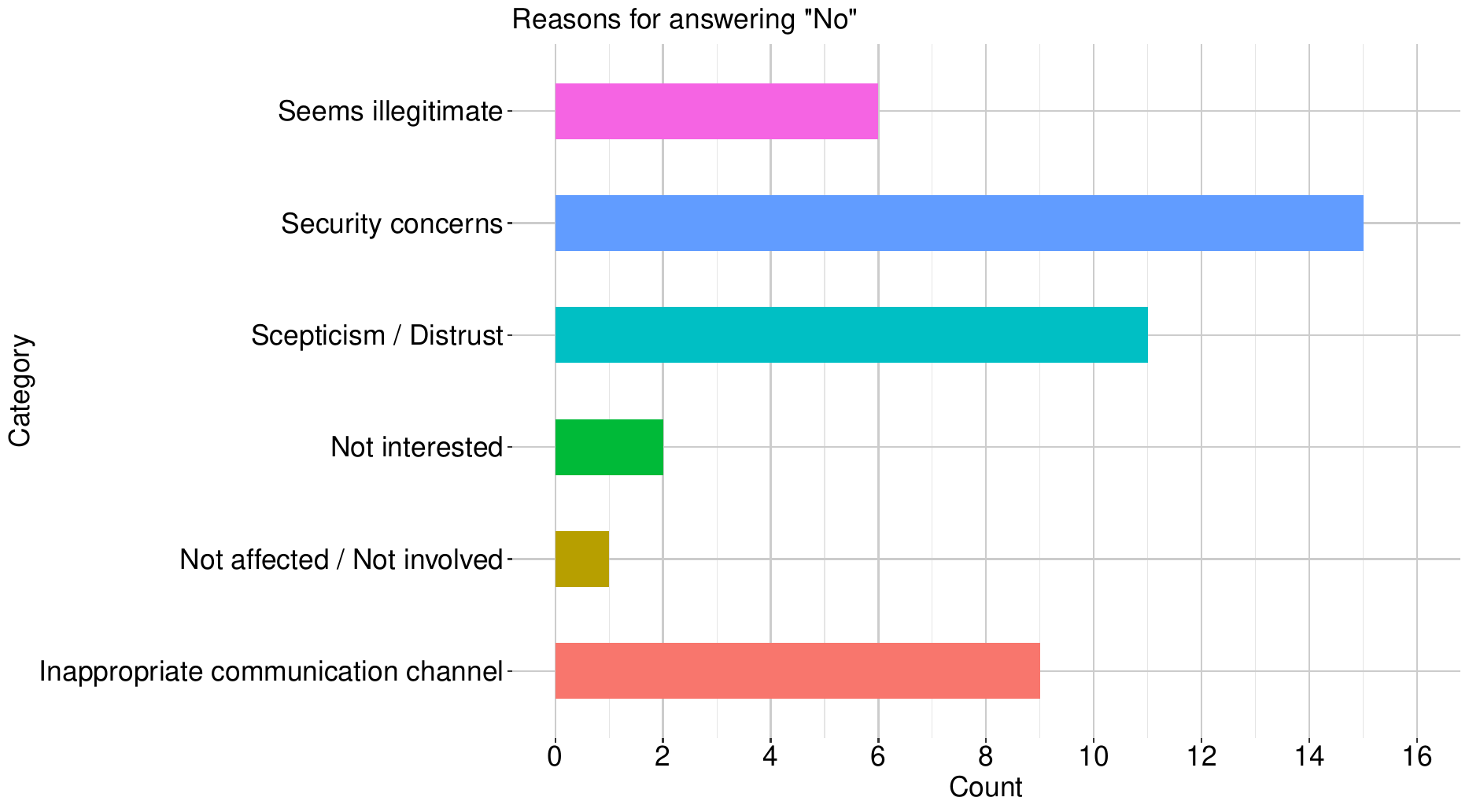}
    \caption{Reasons for answering ``No'' in all stimuli using fear.}
    \label{fig:fear all no}
\end{figure}

\begin{figure}
    \centering
    \includegraphics[width=0.7\linewidth]{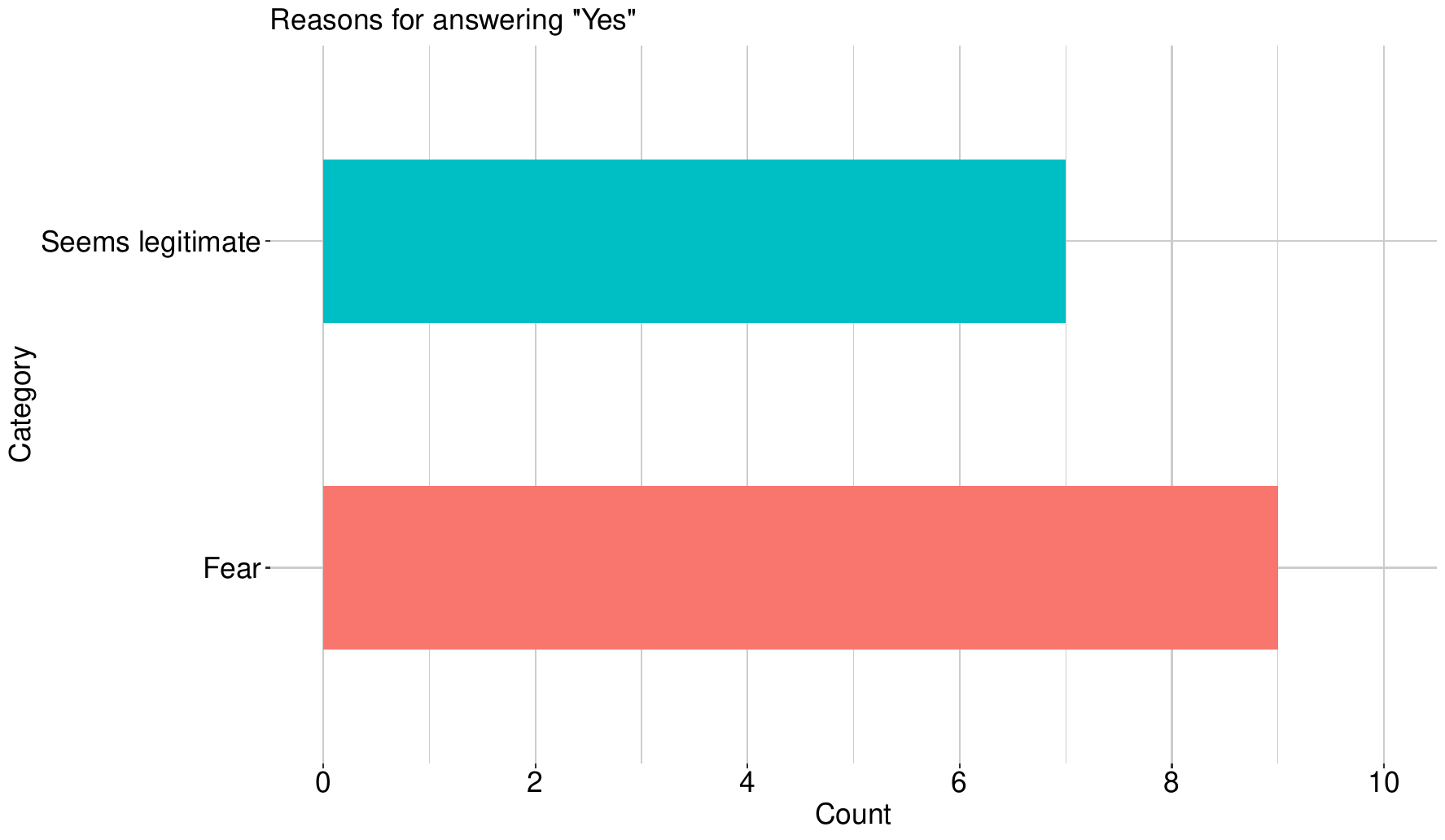}
    \caption{Reasons for answering ``Yes'' in all stimuli using fear.}
    \label{fig:fear all yes}
\end{figure}

Phishing and spear-phishing were the most effective \ac{SE} attack types for fear, each having had a success rate of 41.6\,\%. Smishing followed with 25\,\%, pop-ups with 16.6\,\% and vishing with 8.3\,\%. In total, fear had a success rate of 26.6\,\% across all five scenarios. The most prevalent reason for answering ``Yes'' was fear (\textit{``PayPal has access to my sensitive information''}, \textit{``It concerns my finances''}, see Figure~\ref{fig:fear all yes}). This category constitutes 56.2\,\% of the reasons for answering ``Yes''. Figure~\ref{fig:fear all no} shows that the category ``Security concerns'' leads the reasons for not interacting with the stimuli (34\,\% of the reasons for answering ``No''). Examples from ``Security concerns'' are \textit{``I was asked for very sensitive information''} and \textit{``It would make me feel uncertain, but I would check my PayPal app''}.

\subsubsection{Greed}
The most effective SE attack types for greed were spear-phishing (75\,\% success rate) and phishing (66.6\,\% success rate). Both attacks had more ``Yes'' than ``No'' answers. Pop-ups followed with a 33.3\,\% success rate, smishing with 16.6\,\%, and vishing with 8.3\,\%. The total Yes-rate of greed in all five attacks was 40\,\%. The most common reason for answering ``Yes'' to the greed stimuli was ``Financial motivation'' (54.1\,\% contribution to reasons for ``Yes''). Examples from ``Financial motivation'' are \textit{``I like discounts''} and \textit{``There are benefits for me''}. The most prevalent reason for answering ``No'' to greed stimuli was ``Scepticism / Distrust'', which occurs strongly in vishing, pop-up, and smishing attacks. An exemplary answer from this category was \textit{``The pretext of being selected for an exclusive access to deals seems odd to me''}. 

\subsubsection{Trust}
Trust was most successful in spear-phishing (66.6\,\% success) and phishing (58.3\,\% success). The ``Yes'' answers were more than ``No'' for these two scenarios. Pop-ups and smishing each had a success rate of 16.6\,\%, and vishing 8.3\,\%. The overall success rate of trust in the five \ac{SE} types was 33.3\,\%. Reasons for answering ``Yes'' were mostly in the category ``Seems legitimate'' (65\,\% of Yes-vote reasons), followed by ``Personal relevance / Relationship'' (35\,\% contribution to Yes-votes). Answers from the category ``Seems legitimate'' were \textit{``Looks like a normal email from Netflix''} and \textit{''Looks trustworthy''}. An example from ``Personal relevance / Relationship'' is \textit{``Because I have a Netflix account.''} The most common explanation for answering ``No'' to any stimuli using trust was ``Scepticism / Distrust'' (42.5\,\% contribution to No-votes).

\subsection{Within-SE Design}

\subsubsection{Vishing}

Vishing had a total success rate of 6.6\,\% with the \acp{PF} authority, fear, greed, and trust each having one out of twelve ``Yes'' answers (8.3\,\% success rate each). Vishing using curiosity had no success at all. Most reasons for answering ``No'' to any of the vishing stimuli were from the category ``Scepticism / Distrust'', followed by ``Inappropriate communication channel''. An answer from ``Scepticism / Distrust'' was \textit{``I generally don't trust phone calls''}. An example from ``Inappropriate communication channel'': \textit{``This should be sent as an email.''}

\subsubsection{Phishing}
\begin{figure}
    \centering
    \includegraphics[width=0.94\linewidth]{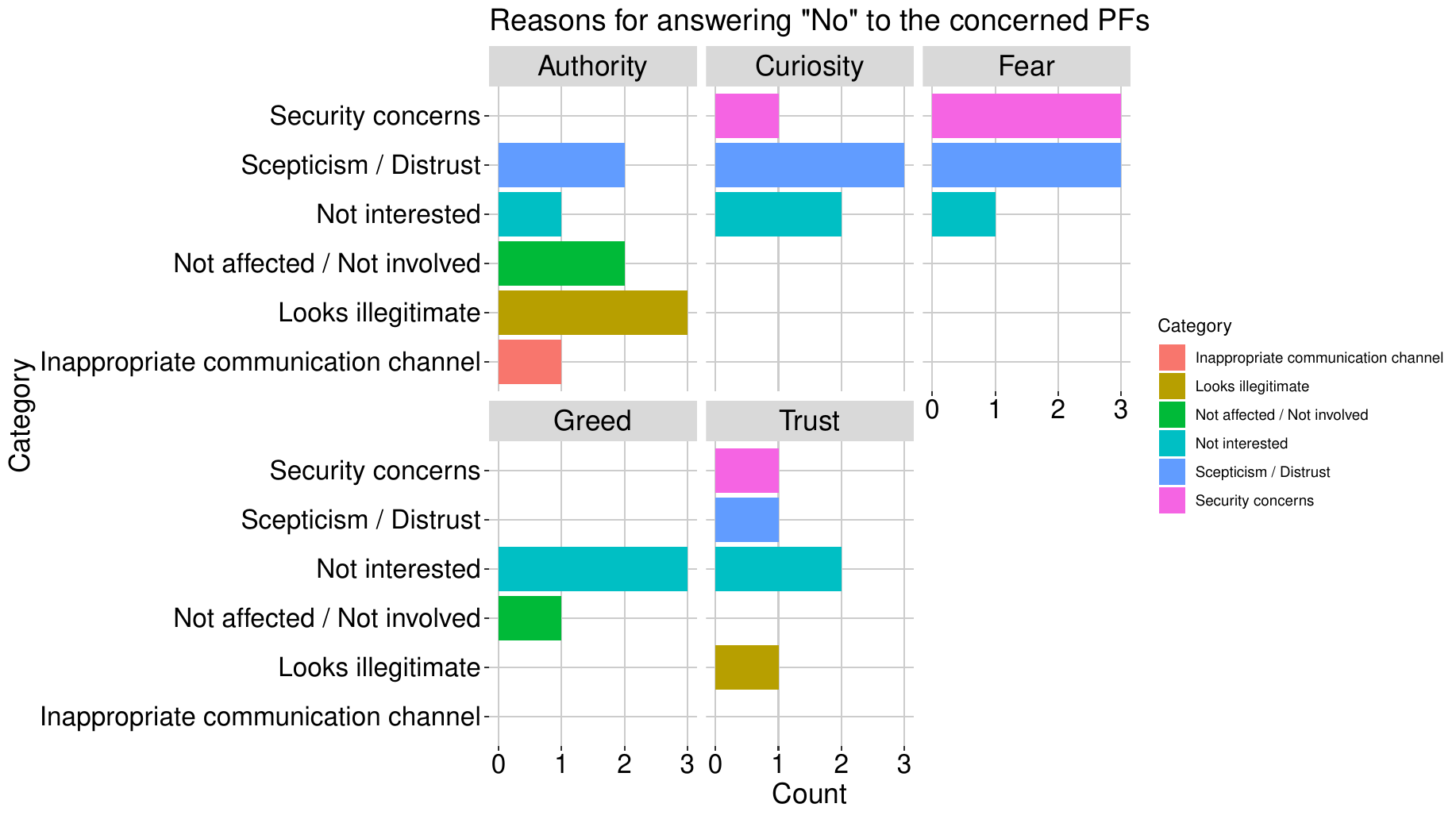}
    \caption{Reasons for answering ``No'' to each concerned PF in phishing attacks.}
    \label{fig:phishing no vote reasons for each pf}
\end{figure}

The overall success rate of phishing was 48.3\,\%. Phishing was most successful by using greed (66.6\,\% success), followed by trust (58.3\,\% success), and curiosity (50\,\% success). Fear in phishing had a success rate of 41.6\,\% and authority 25\,\%. The most common reason for answering ``Yes'' to any phishing stimulus was ``Emotional response''. This category contributes 58.6\,\% to the reasons for Yes-votes. Answers from this category included \textit{``Out of curiosity''}, \textit{``I like discounts''}, and \textit{``PayPal has access to my sensitive information.''} The most frequently mentioned reasons for answering ``No'' to any phishing stimulus were ``Scepticism / Distrust'' and ``Not interested''. Answers from the category ``Scepticism / Distrust'' included \textit{``There is no direct form of address, like `Good day Mr. Doe'\,''} and \textit{``There are many fake emails that look like this''}. Figure~\ref{fig:phishing no vote reasons for each pf} demonstrates the reasons for answering ``No'' 
and shows that the category ``Looks illegitimate'' occurred this often only in authority. 

\vspace{-0.5cm}

\subsubsection{Spear-Phishing}
\begin{figure}
    \centering
    \includegraphics[width=0.8\linewidth]{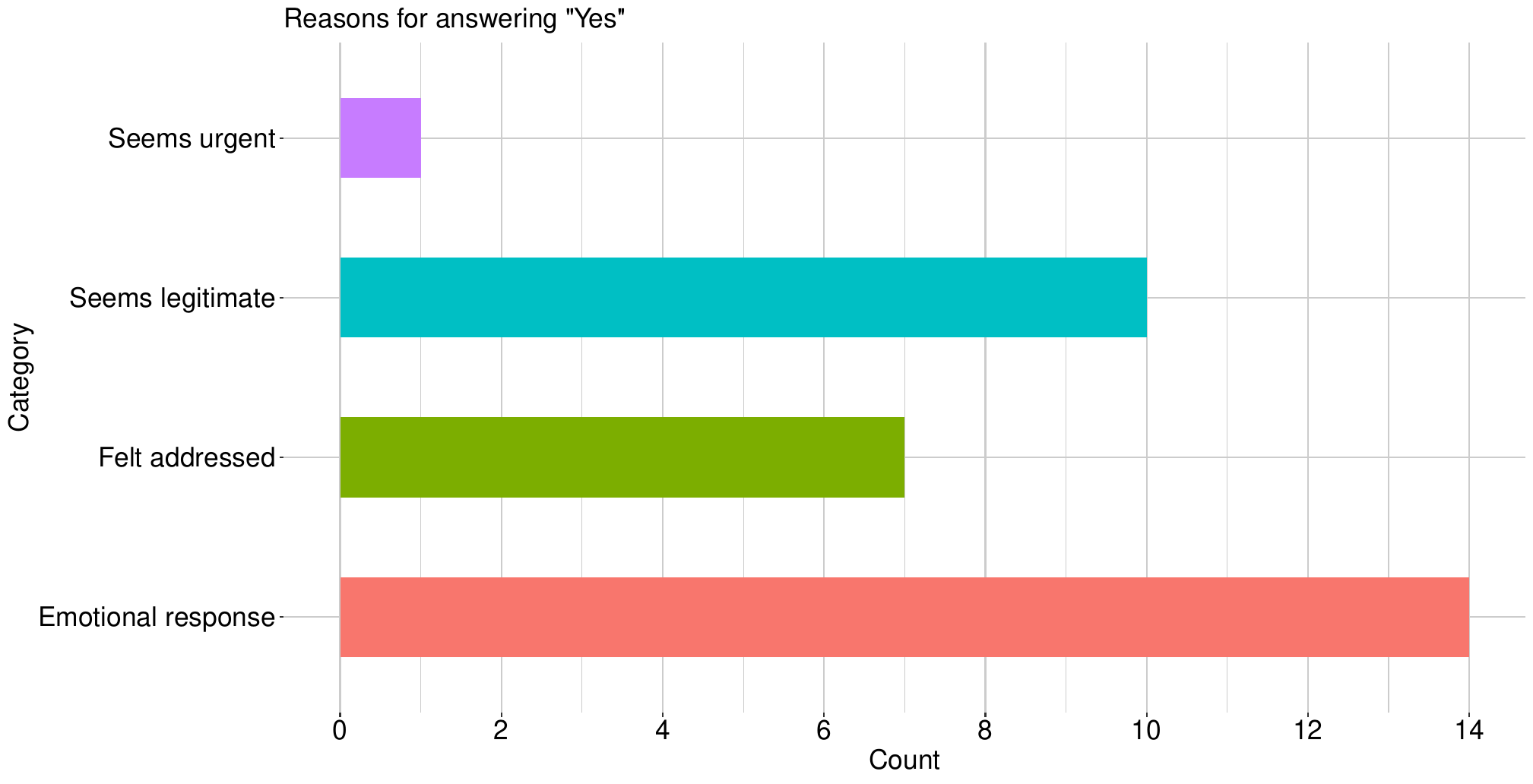}
    \caption{Reasons for answering ``Yes'' to all stimuli using spear-phishing.}
    \label{fig:spear phishing yes reasons total}
\end{figure}

Spear-phishing was most successful by using the \acp{PF} greed (75\,\% success), trust (66.6\,\% success), and curiosity (50\,\% success). Spear-phi\-shing using fear had a success rate of 41.6\,\% and 33.3\,\% for authority. The overall success rate of spear-phishing constitutes 53.3\,\%. As seen in Figure~\ref{fig:spear phishing yes reasons total}, ``Emotional response'' was the most frequent reason for answering ``Yes'' to any spear-phishing stimulus (43.7\,\% contribution to the reasons for yes-votes). Answers from this category included \textit{``It concerns my finances''} and \textit{``I like discounts.''} The category ``Seems legitimate'' follows, and after that comes the category ``Felt addressed''. The latter category included answers such as \textit{``I was addressed directly by my name and would therefore click on the link without much thought''} and \textit{``My account details, such as my name, are provided.''}

\subsubsection{Pop-Up}
The total success rate of pop-up attacks was 15\,\%, with greed in pop-ups being the most useful \ac{PF} (33.3\,\% success), followed by trust and fear (each 16.6\,\% success). Curiosity and authority had success rates lower than 10\,\%. ``Emotional response'' was the most common reason for answering ``Yes''. The most common reason for answering ``No'' was ``Scepticism / Distrust'', followed by ``Inappropriate communication channel''. Answers from ``Scepticism / Distrust'' included \textit{``I don't trust pop-up windows''}, and exemplarily answers from ``Inappropriate communication channel'' is \textit{``I would get notified by my app or emails, not through a pop-up.''}

\subsubsection{Smishing}
Smishing was most successful by using fear with an effectiveness rate of 25\,\%. The overall success rate of smishing was 13.3\,\%. The most common reason for answering ``No'' to any smishing stimulus was ``Inappropriate communication channel''. Answers from that category included \textit{``I'd rather react to an email- an SMS seems strange''} and \textit{``Google wouldn't send me an SMS.''} 

\subsection{Between SE/PF-Design}
\begin{sidewaysfigure}
    \centering
    \includegraphics[width=1\linewidth]{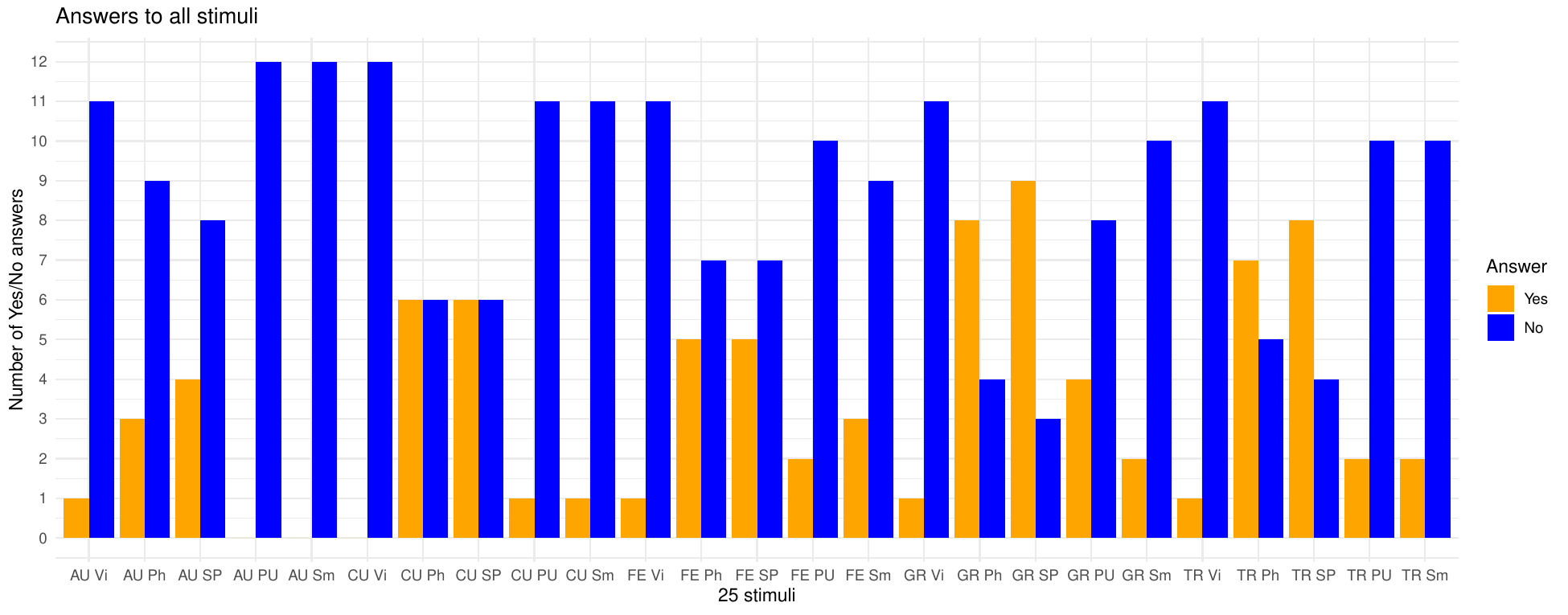}
    \caption{Results from all 25 stimuli.}
    \label{fig:all 25 results}
\end{sidewaysfigure}

This section presents the results for all 25 stimuli, alongside the most and least successful combinations of \ac{PF} and \ac{SE} attack types (see Figure~\ref{fig:all 25 results}). Due to the amount of stimuli, the names of the combinations of \acp{PF} and \ac{SE} attacks will be abbreviated. The abbreviation is as follows:
\begin{multicols}{2}
\begin{itemize}
    \item Authority = AU
    \item Curiosity = CU
    \item Fear = FE
    \item Greed = GR
    \item Trust = TR
    \item Vishing = Vi
    \item Phishing = Ph
    \item Spear-phishing = SP
    \item Pop-Up = PU
    \item Smishing = Sm
\end{itemize}
\end{multicols}

The most successful combination of \ac{PF} and \ac{SE} attack type was spear-phishing using greed, with a success rate of 75\,\%, followed by phishing using greed and spear-phishing using trust (each 66.6\,\% success rate). The third most successful combination was phishing using trust (58.3\,\% success). The least successful combinations were pop-ups using authority, smishing using authority, and vishing using curiosity, each having had no ``Yes'' answers at all. Overall, the most successful \ac{SE} attack type was spear-phishing (53.3\,\% success), followed by phishing (48.3\,\% success). The three most effective \acp{PF} were greed (40\,\% success), trust (33.3\,\% success), and fear (26.6\,\% success). Figure~\ref{fig:all 25 results} shows the results for all 25 stimuli.

\subsection{Self-Rated Digital Risk Awareness}

\begin{figure}
    \centering
    \includegraphics[width=1\linewidth]{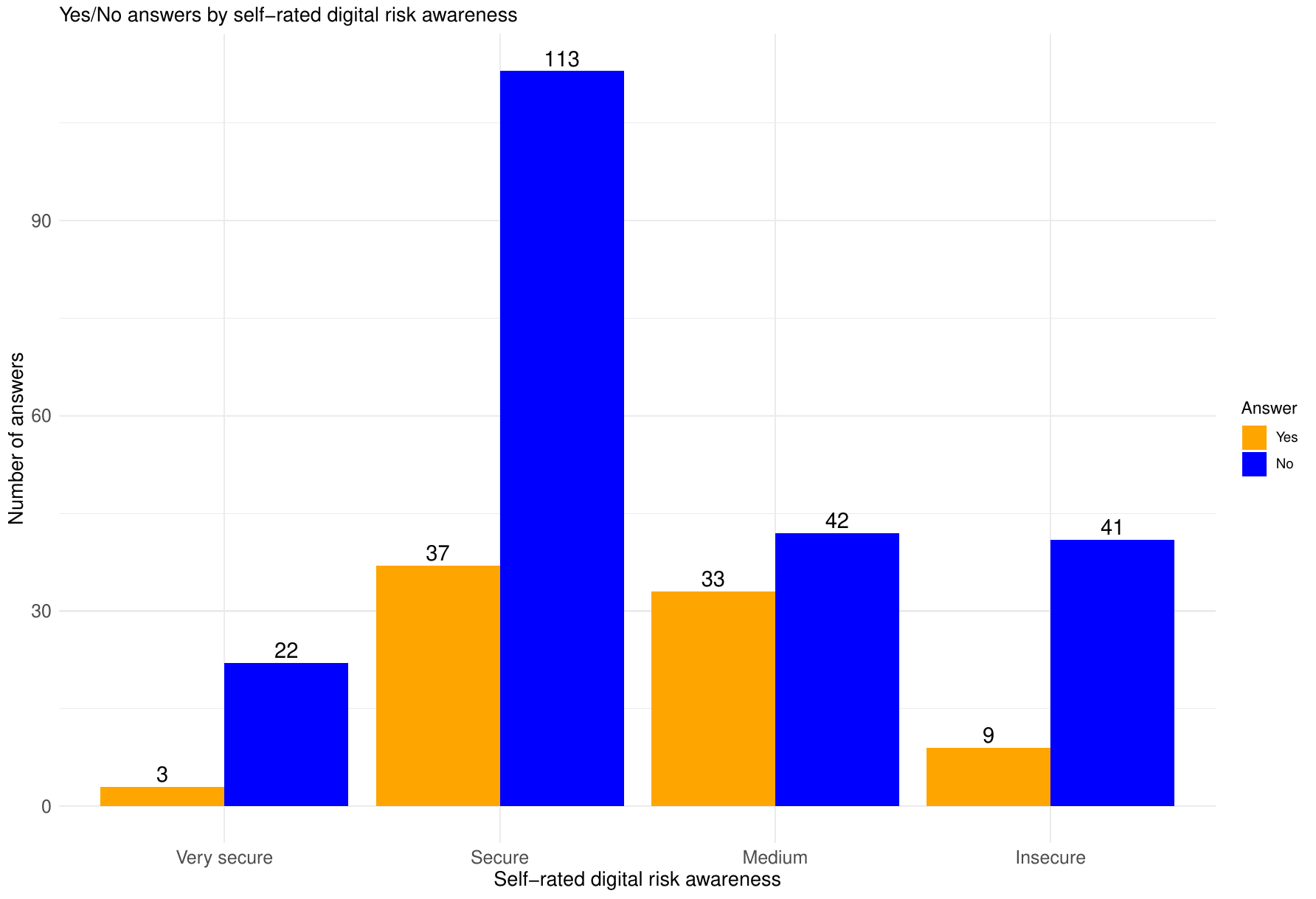}
    \caption{Yes/No answers by self-rated digital risk awareness}
    \label{fig:self-rated digital risk yes/no}
\end{figure}

Figure \ref{fig:self-rated digital risk yes/no} shows the Yes/No answers from participants grouped by their self-rated digital risk awareness. Participants who rated themselves as secure in dealing with digital risks had the highest number of yes-votes (37 yes-votes), followed by the medium group (33 yes-votes). Participants who rated themselves as very secure in dealing with digital risks had the lowest number of yes-votes (3 votes), followed by the insecure group (9 votes).


\section{Discussion}
\label{sec:discussion}

\subsection{Within-PF Design}

\subsubsection{Authority}
In the laboratory study, authority was not particularly successful in any of the five attack scenarios, with a total yes-vote of 13.3\,\%. This is contrary to results from previous studies, such as Butavicius et. al. \cite{butaviciusBreachingHumanFirewall2016}, where authority was the most persuasive strategy and made phishing harder to detect. Modic and Lea \cite{modicScamCompliancePsychology2013} also identified that four key factors, including authority, reliably predict scam compliance. The most frequent reason for answering ``No'' to any authority stimulus was ``Inappropriate communication channel'', followed by ``Seems illegitimate'' (\textit{``It looks fake''}), and ``Scepticism / Distrust''. Besides ``Inappropriate communication channel'', the category ``Seems illegitimate'' is important to consider, since it did not occur as much in the No-vote reasons of other PFs (see Figure~\ref{fig:phishing no vote reasons for each pf} for authority in phishing). Therefore, the limited success of authority may be attributed to two factors: the message content was not appropriate for the used communication channels, and the design/scenario of authority in the emails, SMS, and vishing audio was not convincing enough.

\subsubsection{Curiosity}
Curiosity had a total Yes-vote of 23.3\,\% and showed that the most frequently stated reason for answering ``No'' was within ``Scepticism / Distrust'', followed by ``Not interested''. This indicates that the lower Yes-rate could be improved by using more examples that are applicable to more participants, since some may not use Google Drive or are not interested enough in it. The most frequently occurring reason for answering ``Yes'' was ``Curiosity / Interest'', which shows that the effect of curiosity, however, did work.

\subsubsection{Fear}
In previous literature, such as Longtchi et al. \cite{longtchiInternetBasedSocialEngineering2024a}, fear and greed were the most targeted emotional \acp{PF}. In this laboratory study, fear had a total Yes-vote of 26.6\,\%. Figure~\ref{fig:fear all yes} and Figure~\ref{fig:fear all no} show the reasons for interacting or not interacting with any fear stimuli. The most prevalent reason for answering ``Yes'' was fear (\textit{``It concerns my account and finances''}). The most often stated reason for answering ``No'' to any fear stimulus was ``Security concerns'' (\textit{``It concerns my finances''}), which represents fear operating in the opposite direction. Instead of prompting interaction, fear of losing control over sensitive or financial data prevented participants from engaging with the stimulus. This shows that fear was a dominant factor influencing participants' decisions to interact or not to interact with the stimuli. The relatively low Yes-rate for fear can therefore be explained by its counter effect that prevented participants from interacting.

\subsubsection{Greed}
Atkins and Huang \cite{atkinsStudySocialEngineering2013} found that advance-fee scams emphasized potential monetary gain, while Longtchi et al. \cite{longtchiCharacterizingEvolutionPsychological2025} identified that fear and greed were the most targeted emotional \acp{PF}. This aligns with our study's results, where greed was successful (40\,\% success), especially in spear-phishing (75\,\% success) and phishing attacks (66.6\,\% success). The most common reason for answering ``Yes'' to any greed stimulus was ``Financial motivation'', which supports the effectiveness of greed in the respective stimuli. The category ``Seems legitimate'' follows after, which shows that the design of the stimuli was convincing.

\subsubsection{Trust} 
Alseadoon et al. \cite{alseadoonWhatInfluenceUsers2015}  found that trust increased compliance, while Longtchi et al. \cite{longtchiInternetBasedSocialEngineering2024a} discovered that trust and negligence were the most exploited \ac{PID} \acp{PF}. In this laboratory study, trust had a success rate of 33.3\,\% and was most successful in spear-phishing (66.6\,\% success) and phishing (58.3\,\% success), which aligns with previous literature. Trust in smishing, pop-ups and vishing however, was not as successful (pop-ups and smishing each had a success rate of 16.6 \%, and vishing 8.3 \%). Reasons for not interacting with the trust in smishing stimulus were mostly within the category ``Scepticism / Distrust'', along with trust in vishing and pop-ups. This suggests that these attack types raise more scepticism in general. The most common reason for interacting with any trust stimulus was ``Seems legitimate'' (65\,\% contribution to reasons), which in this case translates to the \ac{PF} trust and demonstrates that trust was accomplished.

\subsection{Within-SE Design}

\subsubsection{Vishing}
Previous literature pointed to the steadily growing use of vishing in 2023, with a 260\,\% increase compared to 2022 Q4 \cite{APWG2023Q4}. In 2024 Q1, phone-based phishing continued to grow steadily, with fraudulent phone numbers making up over 20\,\% of fraud-related assets \cite{apwg}. This aligns with vishing's low success rate in this study (6.6\,\%). Due to the significant rise in vishing attacks, individuals' exposure to these calls has increased, which has heightened their distrust towards calls. The most common reason for answering ``No'' to any vishing stimulus was ``Scepticism / Distrust''.

\subsubsection{Phishing}
Phishing via email continues to be one of the most damaging \ac{SE} attacks \cite{APWG2023Q4,longtchiInternetBasedSocialEngineering2024a}, which is consistent with this study's results (phishing's total success: 48.3\,\%). Previous studies found that authority and trust were among the most persuasive \acp{PF}\cite{butaviciusBreachingHumanFirewall2016,alseadoonWhatInfluenceUsers2015} in phishing emails. Contrary to \cite{butaviciusBreachingHumanFirewall2016}, the results of this study showed that authority had the lowest success rate. An explanation for that is seen in Figure~\ref{fig:phishing no vote reasons for each pf}; the most common reason for answering ``No'' to authority in phishing was ``Looks illegitimate'', which did not occur this often in any of the other four stimuli. This leads to the conclusion that the design of authority in the phishing email was not convincing and led to suspicion. Instead, the most persuasive \acp{PF} for phishing were greed (66.6\,\% success), trust (58.3\,\% success), and curiosity (50\,\% success). This aligns with \cite{alseadoonWhatInfluenceUsers2015}, where trust was among the most successful \acp{PF}. The main reason for Yes-votes was ``Emotional Response'', which demonstrates that the \acp{PF} were implemented successfully.

\subsubsection{Spear-Phishing}
Previous research found that authority was the most effective \ac{PF} for especially spear-phishing emails \cite{butaviciusBreachingHumanFirewall2016}, which is contrary to this study's results, where greed was the most persuasive \ac{PF} for spear-phishing (75\,\% success). As discussed before, authority was not particularly successful in any of the five attacks. The most common reason for No-votes to authority in spear-phishing was ``Not interested'', which did not occur as much in the other \acp{PF}. This suggests that the chosen scenario was responsible for the low success rate, rather than the \ac{PF} authority itself. The most prevalent reason for answering ``Yes'' to any spear-phishing stimulus was ``Emotional Response'', which shows that the \acp{PF} were successfully implemented. The category ``Felt addressed'' was the third most common reason for interacting with any spear-phishing stimulus, which demonstrates that the direct address to recipients was noticed and effective.

\subsubsection{Pop-Up}
Pop-ups were not particularly successful, with an overall success rate of 15\,\%. The most common reason for answering ``No'' was ``Scepticism / Distrust'', followed by ``Inappropriate communication channel''. This suggests that the content was not suitable for pop-ups and that pop-ups seem to be less trustworthy in general.

\subsubsection{Smishing}
Blancafor et al. \cite{inproceedings} found that trust was the most important \ac{PF} for smishing attacks, while Rahman et al. \cite{rahman2022usersreallyrespondsmishing} found that fear and greed had little difference in their effectiveness in smishing.
Our study showed that smishing had the lowest overall success rate (13.3\,\%), with fear being the most effective \ac{PF} (25\,\% success), followed by greed and trust (both 16.6\,\% success).
Overall, smishing was less successful than other attack types because participants viewed SMS as an inappropriate communication channel, while an illegitimate looking design and scepticism also reduced interaction. 

\subsection{Self-Rated Digital Risk Awareness}
Halevi et al. found in one study that self-perceived risk did not predict actual behaviour regarding phishing attacks \cite{haleviPhishingPersonalityTraits2013}, and that participants generally underestimated their own risk \cite{haleviSpearPhishingWildRealWorld2015}. Figure~\ref{fig:self-rated digital risk yes/no} shows that participants who rated themselves as secure and medium in handling digital risks actually overestimated themselves or underestimated their own risk, which aligns with \cite{haleviPhishingPersonalityTraits2013} and \cite{haleviSpearPhishingWildRealWorld2015}. Participants who rated themselves as very secure or insecure were the most cautious, either due to having more knowledge about digital risks or feeling insecure.

 \section{Conclusion} 
\label{sec:conclusion}

Human factors are being exploited in \ac{SE} attacks, with attackers starting to focus on other channels than email. To develop countermeasures like awareness campaigns, it is important to know how effective these attacks are and which \acp{PF} and \acp{PT} are being most successful. In our study, we exposed the participants with different SE attacks and \acp{PF}/\acp{PT}. The exploratory results showed that the most effective \ac{SE} attack type for authority, greed, and trust was spear-phishing. The most effective \ac{SE} attack types for curiosity and fear were phishing and spear-phishing. Authority was not particularly successful, with a success rate of 13.3\,\% in comparison with greed (40\,\%) and trust (33.3\,\%) that had the highest success rates among the five \acp{PF}. The results also show that vishing (8.3\,\% success rate) was relatively unsuccessful. Overall, the most successful attack types were spear-phishing (53.3\,\% success rate) and phishing (48.3\,\% success rate). The most successful combinations were spear-phishing with greed (75\,\% success rate), phishing with greed, and spear-phishing with trust (both with 66.6\,\% success rate). In future work, we want to include more participants, \ac{SE} attacks, and \acp{PF}/\acp{PT}, as well as more variations of the combinations to validate and extend our results.

\vspace{-0.1cm}
\subsubsection*{Declaration:} The authors have no competing interests to declare that are relevant to the content of this article.

%
%
%
\begin{acronym}
\acro{PF}{psychological factor}
\acro{SE}{social engineering}
\acro{vishing}{phone-based voice phishing}
\acro{smishing}{SMS-based phishing}
\acro{PT}{psychological technique}
\acro{OSN}{online social network}
\acro{PID}{personality and individual difference}
\end{acronym}

\bibliographystyle{splncs04}
\bibliography{workshop}

@inproceedings{alseadoonWhatInfluenceUsers2015,
  title = {{What Is the Influence of Users’ Characteristics on Their Ability to Detect Phishing Emails?}},
  booktitle = {Advanced Computer and Communication Engineering Technology},
  author = {Alseadoon, Ibrahim and Othman, M. F. I. and Chan, Taizan},
  editor = {Sulaiman, Hamzah Asyrani and Othman, Mohd Azlishah and Othman, Mohd Fairuz Iskandar and Rahim, Yahaya Abd and Pee, Naim Che},
  year = {2015},
  pages = {949--962},
  publisher = {Springer International Publishing},
  location = {Cham},
  doi = {10.1007/978-3-319-07674-4_89}
}

@article{atkinsStudySocialEngineering2013,
  title = {A {{Study}} of {{Social Engineering}} in {{Online Frauds}}},
  author = {Atkins, Brandon and Huang, Wilson},
  date = {2013-01-01},
  year = {2023},
  journal = {Open Journal of Social Sciences},
  shortjournal = {Open Journal of Social Sciences},
  volume = {01},
  pages = {23--32},
  doi = {10.4236/jss.2013.13004}
}

@article{bulleeAnatomySocialEngineering2018,
  title = {On the Anatomy of Social Engineering Attacks: {{A}} Literature-Based Dissection of Successful Attacks},
  shorttitle = {On the Anatomy of Social Engineering Attacks},
  author = {Bullée, Jan-Willem and Montoya, Lorena and Pieters, Wolter and Junger, Marianne and Hartel, Pieter},
  year = {2018},
  journal = {Journal of Investigative Psychology and Offender Profiling},
  volume = {15},
  number = {1},
  pages = {20--45},
  publisher = {Wiley},
  issn = {1544-4759},
  doi = {10.1002/jip.1482},
  url = {https://research.utwente.nl/en/publications/on-the-anatomy-of-social-engineering-attacks-a-literature-based-d},
  urldate = {2025-09-03},
  langid = {english}
}

@misc{butaviciusBreachingHumanFirewall2016,
  title = {Breaching the {{Human Firewall}}: {{Social}} Engineering in {{Phishing}} and {{Spear-Phishing Emails}}},
  author = {Butavicius, Marcus and Parsons, Kathryn and Pattinson, Malcolm and McCormac, Agata},
  year = {2016},
  doi = {10.48550/arXiv.1606.00887},
  howpublished = {\url{url={https://arxiv.org/abs/1606.00887}, }}
}

@misc{caniphishSecurityAwarenessTraining,
  title = {{Security Awareness Training | Phishing Simulation | CanIPhish}},
  author = {{canIPhish}},
  howpublished = {\url{https://caniphish.com/User/WebPhishing}},
  urldate = {2025-11-11}
}

@article{cialdiniSciencePersuasion2001,
  title = {The {{Science}} of {{Persuasion}}},
  author = {Cialdini, Robert B.},
  year = {2001},
  journal = {Scientific American},
  volume = {284},
  number = {2},
  eprint = {26059056},
  eprinttype = {jstor},
  pages = {76--81},
  publisher = {Scientific American, a division of Nature America, Inc.},
  issn = {0036-8733},
  url = {https://www.jstor.org/stable/26059056},
  urldate = {2025-08-18}
}

@misc{elevenlabsAIVoiceGenerator,
  title = {{AI Voice Generator \& Text to Speech}},
  author = {{ElevenLabs}},
  howpublished = {\url{https://elevenlabs.io}},
  year = {2026},
  urldate = {2025-11-19}
}

@misc{haleviPhishingPersonalityTraits2013,
  title = {Phishing, {{Personality Traits}} and {{Facebook}}},
  author = {Halevi, Tzipora and Lewis, Jim and Memon, Nasir},
  year = {2013},
      eprint={1301.7643},
      archivePrefix={arXiv},
      primaryClass={cs.HC},
      howpublished={\url{https://arxiv.org/abs/1301.7643}}
}

@article{haleviSpearPhishingWildRealWorld2015,
  title = {Spear-{{Phishing}} in the {{Wild}}: {{A Real-World Study}} of {{Personality}}, {{Phishing Self-Efficacy}} and {{Vulnerability}} to {{Spear-Phishing Attacks}}},
  shorttitle = {Spear-{{Phishing}} in the {{Wild}}},
  author = {Halevi, Tzipora and Memon, Nasir and Nov, Oded},
  year = {2015},
  journal = {SSRN Electronic Journal},
  shortjournal = {SSRN Journal},
  issn = {1556-5068},
  doi = {10.2139/ssrn.2544742},
  url = {http://www.ssrn.com/abstract=2544742},
  urldate = {2025-09-03},
  langid = {english}
}

@misc{hessFakeTextMessage,
  title = {{Fake Text Message | Make Fake Text Conversation}},
  author = {Hess, Dillon},
  howpublished = {\url{https://ifaketextmessage.com}},
  year = {2023},
  urldate = {2025-11-16},
  langid = {english}
}

@inproceedings{longtchiCharacterizingEvolutionPsychological2025,
  title = {Characterizing the~{{Evolution}} of~{{Psychological Factors Exploited}} by~{{Malicious Emails}}},
  booktitle = {Science of {{Cyber Security}}},
  author = {Longtchi, Theodore and Xu, Shouhuai},
  editor = {Zhao, Jun and Meng, Weizhi},
  year = {2025},
  pages = {158--178},
  publisher = {Springer Nature},
  location = {Singapore},
  doi = {10.1007/978-981-96-2417-1_9},
  url = {https://link.springer.com/chapter/10.1007/978-981-96-2417-1_9},
  isbn = {978-981-96-2417-1},
  langid = {english}
}

@article{longtchiInternetBasedSocialEngineering2024a,
  title = {{Internet-Based Social Engineering Psychology, Attacks, and Defenses: A Survey}},
  author = {Longtchi, Theodore Tangie and Rodriguez, Rosana Montañez and Al-Shawaf, Laith and Atyabi, Adham and Xu, Shouhuai},
  year = {2024},
  journal = {Proceedings of the IEEE},
  volume = {112},
  number = {3},
  pages = {210--246},
  issn = {0018-9219, 1558-2256},
  doi = {10.1109/JPROC.2024.3379855},
  url = {https://ieeexplore.ieee.org/document/10493072/},
  urldate = {2025-08-18}
}

@article{modicScamCompliancePsychology2013,
  title = {Scam {{Compliance}} and the {{Psychology}} of {{Persuasion}}},
  author = {Modic, David and Lea, Stephen E. G.},
  year = {2013},
  journal = {SSRN Electronic Journal},
  shortjournal = {SSRN Journal},
  issn = {1556-5068},
  doi = {10.2139/ssrn.2364464},
  url = {http://www.ssrn.com/abstract=2364464},
  urldate = {2025-10-29},
  langid = {english}
}

@book{moutonSocialEngineeringAttack2014,
  title = {Social {{Engineering Attack Framework}}},
  author = {Mouton, Francois and Malan, Mercia and Leenen, Louise and Venter, H.s},
  year = {2014},
  journaltitle = {Information Security for South Africa},
  doi = {10.1109/ISSA.2014.6950510},
  publisher = {Scientific Research Publishing}
}

@inproceedings{pozoSocialEngineeringApplication2018,
  title = {Social {{Engineering}}: {{Application}} of {{Psychology}} to {{Information Security}}},
  author = {Pozo, Ivan and Iturralde, Mauricio and Restrepo-Calle, Felipe},
  year = {2018},
  pages = {108--114},
  doi = {10.1109/W-FiCloud.2018.00023},
    booktitle={2018 6th International Conference on Future Internet of Things and Cloud Workshops (FiCloudW)}
}

@article{wallPersonalityProfilesPersuasion2019,
  title = {Personality Profiles and Persuasion: {{An}} Exploratory Study Investigating the Role of the {{Big-5}}, {{Type D}} Personality and the {{Dark Triad}} on Susceptibility to Persuasion},
  shorttitle = {Personality Profiles and Persuasion},
  author = {Wall, Helen J. and Campbell, Claire C. and Kaye, Linda K. and Levy, Andy and Bhullar, Navjot},
  year = {2019},
  journal = {Personality and Individual Differences},
  shortjournal = {Personality and Individual Differences},
  volume = {139},
  pages = {69--76},
  issn = {01918869},
  doi = {10.1016/j.paid.2018.11.003},
  url = {https://linkinghub.elsevier.com/retrieve/pii/S0191886918305865},
  urldate = {2025-09-14},
  langid = {english}
}

@misc{verizon,
author = {{Verizon}},
title = {{2025 Data Breach Investigations Report}},
howpublished = {\url{https://www.verizon.com/business/resources/reports/dbir/}},
year = {2025}
}

@misc{sans,
author = {{SANS Institute}},
title = {{SANS Report Finds Humans Still The Main Attack Vector as 80\% of Organizations Flag Social Engineering as Their Number One Risk}},
howpublished = {\url{https://www.sans.org/press/announcements/security-awareness-report-2025}},
year = {2025}
}

@misc{apwg,
author = {{APWG}},
title = {{Phishing Activity Trends Report 1st Quarter 2024}},
year = {2024},
howpublished = {\url{https://docs.apwg.org/reports/apwg_trends_report_q1_2024.pdf}}
}

@Article{app12126042,
AUTHOR = {Siddiqi, Murtaza Ahmed and Pak, Wooguil and Siddiqi, Moquddam A.},
TITLE = {{A Study on the Psychology of Social Engineering-Based Cyberattacks and Existing Countermeasures}},
JOURNAL = {Applied Sciences},
VOLUME = {12},
YEAR = {2022},
NUMBER = {12},
ARTICLE-NUMBER = {6042},
URL = {https://www.mdpi.com/2076-3417/12/12/6042},
ISSN = {2076-3417},
DOI = {10.3390/app12126042}
}

@article{10.1145/3767320,
author = {Tan Jia Jun, Dennis and Sahban Rafsanjani, Ahmad and Aslam, Saad and Behjati, Mehran},
title = {{Human Factors in Information Security: A Quantitative Study with Technical Solutions to Prevent Social Engineering Attacks}},
year = {2025},
issue_date = {December 2025},
publisher = {Association for Computing Machinery},
address = {New York, NY, USA},
volume = {6},
number = {4},
url = {https://doi.org/10.1145/3767320},
doi = {10.1145/3767320},
journal = {Digital Threats},
month = dec,
articleno = {31},
numpages = {35}
}

@article{10.1145/3635149,
author = {Burda, Pavlo and Allodi, Luca and Zannone, Nicola},
title = {{Cognition in Social Engineering Empirical Research: A Systematic Literature Review}},
year = {2024},
issue_date = {April 2024},
publisher = {Association for Computing Machinery},
address = {New York, NY, USA},
volume = {31},
number = {2},
issn = {1073-0516},
url = {https://doi.org/10.1145/3635149},
doi = {10.1145/3635149},
journal = {ACM Trans. Comput.-Hum. Interact.},
month = jan,
articleno = {19},
numpages = {55}
}

@article{CHRYSANTHOU2024103780,
title = {The anatomy of deception: Measuring technical and human factors of a large-scale phishing campaign},
journal = {Computers \& Security},
volume = {140},
pages = {103780},
year = {2024},
issn = {0167-4048},
doi = {doi.org/10.1016/j.cose.2024.103780},
url = {https://www.sciencedirect.com/science/article/pii/S0167404824000816},
author = {Anargyros Chrysanthou and Yorgos Pantis and Constantinos Patsakis},
}

@article{COBBCLARK201211,
title = {The stability of big-five personality traits},
journal = {Economics Letters},
volume = {115},
number = {1},
pages = {11-15},
year = {2012},
issn = {0165-1765},
doi = {https://doi.org/10.1016/j.econlet.2011.11.015},
url = {https://www.sciencedirect.com/science/article/pii/S0165176511004666},
author = {Deborah A. Cobb-Clark and Stefanie Schurer}
}

@book{chaiken1999dual,
  title={Dual-process theories in social psychology},
  author={Chaiken, Shelly and Trope, Yaacov},
  year={1999},
  publisher={Guilford Press}
}

@techreport{APWG2023Q4,
  author       = {Anti-Phishing Working Group},
  title        = {Phishing Activity Trends Report, 4th Quarter 2023},
  institution  = {Anti-Phishing Working Group},
  year         = {2023},
  month        = {Nov},
  type         = {Technical Report},
  url          = {https://docs.apwg.org/reports/apwg_trends_report_q4_2023.pdf},
  note         = {Zugriff am 19.04.2026}
}

@inproceedings{inproceedings,
author = {Blancaflor, Eric and Alfonso, Adrian and Banganay, Kevin and Cruz, Gabriel and Fernandez, Karen and Santos, Shawn},
year = {2021},
month = {03},
pages = {},
title = {Let's Go Phishing: A Phishing Awareness Campaign Using Smishing, Email Phishing, and Social Media Phishing Tools},
doi = {10.46254/AN11.20211105}
}

@misc{rahman2022usersreallyrespondsmishing,
      title={Users really do respond to smishing}, 
      author={Muhammad Lutfor Rahman and Daniel Timko and Hamid Wali and Ajaya Neupane},
      year={2022},
      eprint={2212.13312},
      archivePrefix={arXiv},
      primaryClass={cs.CR},
      url={https://arxiv.org/abs/2212.13312}, 
}

\end{document}